\documentclass[a4paper,reqno,11pt]{amsart} 
\usepackage{amsthm,amsmath,amssymb}
\usepackage{MnSymbol,bbm} 
\usepackage[T1]{fontenc}
\usepackage[normalem]{ulem}
\usepackage{colortbl}
\usepackage{booktabs,wrapfig}         
\usepackage[linktocpage=true]{hyperref}

\usepackage{caption}
\usepackage{subcaption}

\usepackage{float}
\usepackage{wrapfig}

\usepackage{graphicx}
\usepackage{psfrag}

\usepackage{framed}

\usepackage{color} 

\newcounter{mnotecount}[section]

\newcounter{mymnotecount}[section]

\renewcommand{\Re}{{\mathbb R}}		
\newcommand{\half}{\frac{1}{2}}		

\newcommand{\MM}{\mathcal M}

\newcommand{\Scri}{\mathcal I}

\newcommand{\aM}{\mathbf M}
\newcommand{\ame}{\mathbf g}
\newcommand{\aR}{\mathbf R}

\newcommand{\aT}{\mathbf T}

\newcommand{\Ric}{\text{Ric}}

\newcommand{\HH}{\mathbb H}

\newcommand{\Ein}{\mathcal E}

\newcommand{\norm}{\mathbf u}

\theoremstyle{plain}

\title{Cosmological models and stability} 

\author[L. Andersson]{Lars Andersson}
\thanks{Based on a talk at the conference ``Relativity and Gravitation,
100 Years after Einstein in Prague'', at Charles University,
Prague, June 25-29, 2012}
\email{laan@aei.mpg.de}
\address{Albert Einstein Institute, Am M\"uhlenberg 1, D-14476 Potsdam,
  Germany}

\begin{document}

\date{\today \ {\em File:\jobname{.tex}}}

\begin{abstract}
Principles in the form of heuristic guidelines 
or 
generally accepted dogma play an important
role in the development of physical theories. In particular, 
philosophical considerations and principles figure prominently in the work of 
Albert Einstein. 
As mentioned in
the talk by 
Ji{\v{r}}{\'{\i}} Bi{\v{c}}{\'a}k  
at this conference
Einstein formulated the equivalence principle, an essential step on the road
to general relativity, during his time in Prague
1911-1912. 
In this talk, 
I would like to discuss some aspects of cosmological models. 
As cosmology is an area of physics where ``principles''  such as
the ``cosmological principle'' or the ``Copernican principle'' 
play a prominent role in
motivating the class of models which form part of the current standard model,
I will start by comparing the role of the equivalence  principle to that of
the principles used in cosmology. I will then briefly describe the standard
model of cosmology to give a perspective on some mathematical problems and
conjectures on cosmological models, which are discussed in the later part of
this paper.

\end{abstract}

\maketitle

\begin{quote}
\small \emph{
I would already have concluded my researches about world harmony, had not
Tycho's astronomy so shackled me that I nearly went out of my mind.}

\ \ \hfill Johannes Kepler\footnote{%
Letter to Herwart, quoted in
\cite[p. 127]{caspar:1993kepl.book.....C}}
\end{quote}

\section{Introduction}

As stated by 
Einstein in his paper from 1912 \cite{einstein:1912AnP...343.1059E},
submitted just before his departure from Prague, the equivalence principle is  
\textit{``eine Nat\"urliche Extrapolation einer
  der allgemeinsten Erfahrungss\"atze der Physik''}\footnote{``a natural extrapolation of one of the most
general empirical propositions of physics''},  
and 
can consequently be claimed 
to be exactly valid on all scales. 
Since the equivalence 
principle is
compatible with Einstein's 
relativity principle of 1905 only in the
limit of constant gravitational potential,
accepting the principle of equivalence 
meant that a new 
foundation for the theory of gravitation must be
sought. The challenge of doing so, which Einstein in his 1912 paper poses to
his colleagues: \textit{ 
``Ich m\"ochte alle
Fachgenossen  bitten, sich an diesem wichtigen Problem zu
versuchen!''}, is one that he himself devoted the coming years to, finally
arriving at the 1915 theory of general relativity.

General Relativity  describes the
  universe as a 4-manifold $\aM$ with a metric
$\ame_{\alpha\beta}$ of Lorentzian signature.
The Einstein equations, 
\begin{equation}\label{eq:EFE} 
\aR_{\alpha\beta} - \half \aR \ame_{\alpha\beta} + \Lambda \ame_{\alpha\beta}
= 8\pi G \aT_{\alpha\beta} \,,
\end{equation} 
originally given in
\cite{einstein:1915SPAW.......844E}, 
relate the geometry of spacetime $(\aM, \ame_{\alpha\beta})$ to
matter fields with energy-momentum tensor $\aT_{\alpha\beta}$.
By the correspondence principle, the stress energy tensor $\aT_{\alpha\beta}$
should correspond to the stress energy tensor of a special relativistic
matter model, and in particular be divergence free. For ``ordinary matter'' one
expects $\aT_{\alpha\beta}$ to satisfy energy conditions such as the dominant
energy conditon. 
Here I have included the ``cosmological constant term'' $\Lambda
  \ame_{\alpha\beta}$ in   \eqref{eq:EFE}, which 
was not 
present in the equations given in
\cite{einstein:1915SPAW.......844E}. 
The left hand side of \eqref{eq:EFE},
  where $\aR_{\alpha\beta}$ is the Ricci tensor, $\aR$ is the Ricci scalar
  and $\Lambda$ is a constant, 
  is the most general covariant tensor expression of vanishing divergence, 
depending on 
  $\ame_{\alpha\beta}$ and its derivatives up to second order, and linear in
  second derivatives. Further, its left hand side is 
the most general second order
  Euler-Lagrange equation, 
derived by varying a covariant Lagrange
  density defined in $\ame_{\alpha\beta}$ and its first two derivatives, 
see \cite{lovelock:1969ArRMA..33...54L,lovelock:MR0270285} and references
therein.  
The covariance of the equations of general relativity under spacetime
diffeomorphisms, makes the theory 
compatible with the strong version 
of the equivalence principle.

Since it can be claimed to be exactly valid, the equivalence
principle is subject to empirical tests and there is a long history of
experiments testing various versions of the (weak or strong) 
equivalence principles, see
e.g. \cite{haugan:lammerzahl:2001LNP...562..195H}, see also 
\cite{bicak:ae100prg_talk} in this volume. 
Until the present, 
the equivalence principle has survived all experimental tests,  
and an experiment clearly demonstrating a deviation from the
predictions based on the equivalence principle would necessitate a revision
of the foundations of modern physics.

The arguments of the physicist and philosopher Ernst Mach 
played an important role in the development of Einstein's ideas
leading up to general relativity, including the formulation of the
equivalence principle. 
The fact that 
in general relativity, 
matter influences 
the motion of test particles via its effect on spacetime
curvature means that in contrast to Newtonian gravity, the 
 ``action at a distance'' which was critizised by Mach is not present 
in general
relativity, which hence agrees with the guiding idea which 
Einstein referred to as ``Mach's principle'', 
i.e. loosely speaking the idea that the
distribution of matter in the universe determines local frames of inertia,
see \cite{einstein:1918AnP...360..241E}, see also
\cite{barbour:1990mcr..book...47B}. The role of Mach's
principle in the context of cosmology is discussed in
\cite{bicak:etal:2007PhRvD..76f3501B}.
This
played a central role in Einstein's development of
general relativity, and also in his discussion of general relativistic
cosmology, but it 
appears difficult to formulate 
experimentally testable consequences, cf. \cite{will:1995mpfn.conf..365W},  
although Mach's principle has of course 
been brought up in connection with
``Newton's bucket'' and frame dragging. The book
\cite{barbour:pfister:1995mpfn.conf.....B} gives an excellent overview of
issues related to Mach's principle.
However, the principles 
which are most relevant for the present discussion are 
the hierarchy of ``cosmological
principles'', for example 
the cosmological principle of Einstein and the perfect cosmological principle
of Bondi, Gold and Hoyle. See \cite[\S 2.1]{lahav:suto:2004LRR.....7....8L}
  for an overview of the cosmological principles. 
These principles play a role which is fundamentally different from that
of the equivalence principle, in 
the sense that they do not make predictions which are expected to be exactly
true at all scales. At best, they can be viewed as simplifying assumptions
that enable one to construct testable physical models. 

\begin{wrapfigure}{O}{0.4\textwidth}
\centering
\includegraphics[width=0.38\textwidth]{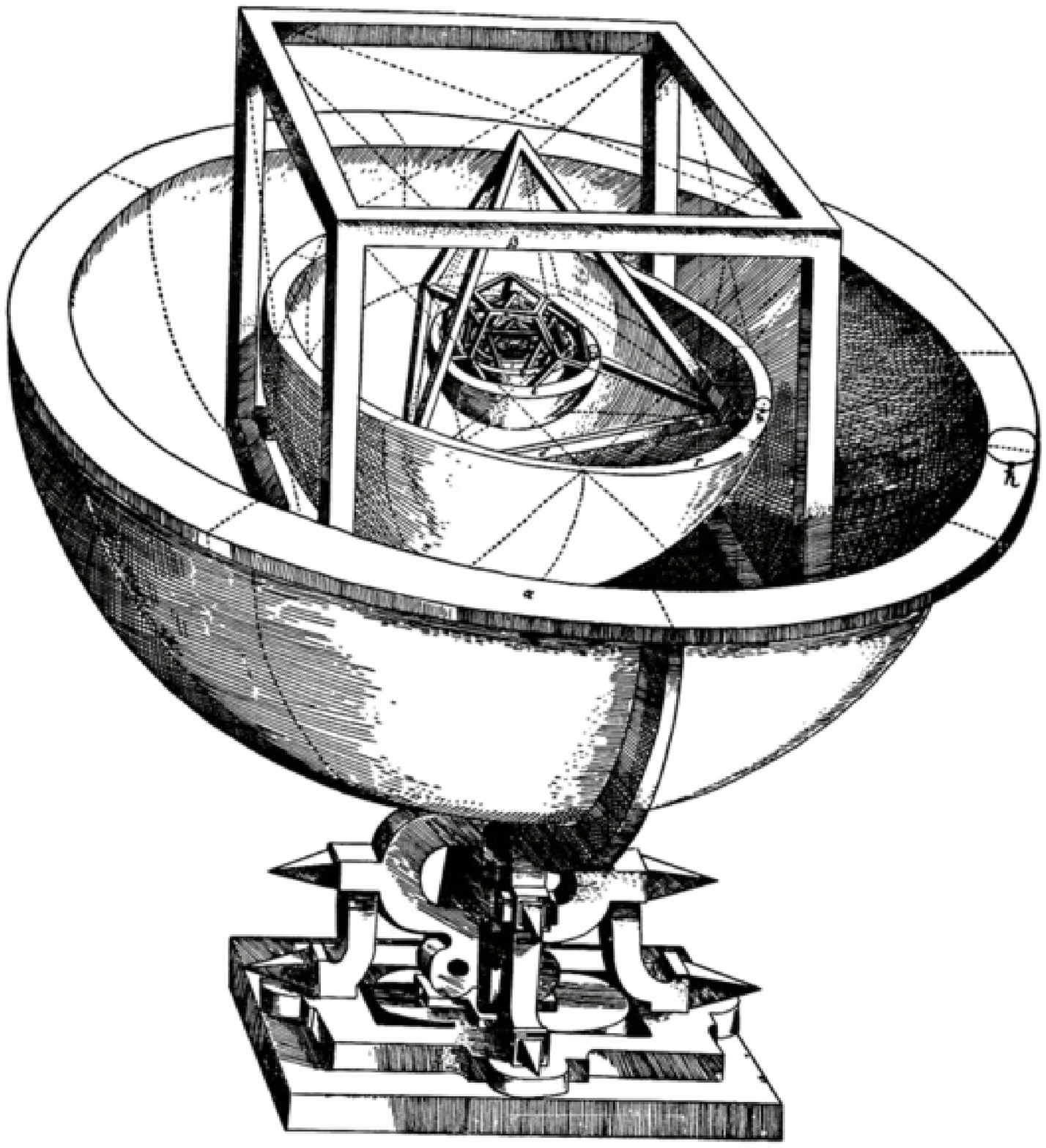}
Kepler's model of the solar system based on Platonic solids, from
  Mysterium Cosmographicum (1596).
\end{wrapfigure} 
The work of Kepler, 
who is perhaps more intimately
connected with Prague than Einstein, provides an
interesting illustration of the relationship between theoretical principle
and observation. In the time of Kepler, the 
world-model of Copernicus had 
placed the sun at the center of the universe 
and described the planets as
moving on circular orbits around it. Not long before his move to Prague in 1600,
Kepler believed himself to have completed the Copernican world-model based on
the mathematical perfection of circles, by adding to it an element of equal
perfection and beauty,
namely the geometry of the Platonic solids, which according to Kepler's
expectations would determine the sizes of the planetary orbits.

Fortunately, 
it was possible for Kepler to use Tycho Brahe's observational data to test
the predictions of his model. To his deep consternation Kepler 
realized that the planets do not, after all,
move on circular orbits. The beautiful principles which had inspired Kepler
to laboriously analyze the observational data of Tycho had to be discarded. 
In analyzing the data, Kepler not only discovered his
three laws of planetary motion but also came close to introducing the notion
of force which became fully clear only through the work of Newton. One could
say that through the work of Kepler and later Newton, one set of ``a priori'' principles
(those of Copernicus and Kepler) were replaced by a model based on the
dynamical laws
of Newtonian gravity.

\subsection{The Cosmological principle} Although 
Newtonian ideas continued to 
dominate physics throughout the 19th century, 
there were well known anomalies of a theoretical as well as
observational nature, and these served as a guide for the developments of the
early 20th century. 
The conflict between the covariance of 
Maxwell theory under the Lorentz group and the more restricted invariance
properties of the Newtonian laws led to the introduction of
special relativity. 
Similarly, as discussed above, the incompatibility of
special relativity and gravitation led to the development of 
general relativity. 
The explanation of the anomalous 
precession of the perihelion of Mercury 
\cite{leverrier:1859AnPar...5....1L}\footnote{From the perspective of the current situation in physics, it is amusing to
recall that 
attempts had been made in the 19th century to explain the obseved precession of
Mercury both by
dark matter \cite{lescarbault:leverrier:1860AnPar...5..394L} 
(the planet Vulcan hypothesis) as well as 
modifications of gravity \cite{hall:1894AJ.....14...49H}.} 
by general relativity
was, together with its new prediction for the 
deflection of light by the sun, confirmed by subsequent observations 
\cite{dyson:etal:1920RSPTA.220..291D}, 
were among the factors which 
led to its rapid acceptance.

Among the main paradoxes of Newtonian physics and world view 
in applications to cosmology were
Olbers' paradox 
and the incompatibility of Newtonian gravity with infinitely extended homogenous
matter distributions, which 
had prevented the construction of a cosmological
model consistent with Newtonian ideas. 
This latter fact, which had been 
elucidated by von Seeliger 
and others, see
\cite{norton:1999ewgr.book..271N} for discussion and references, 
played an
important role in Einstein's reasoning about cosmological models in his 1917
paper \cite{einstein:1917SPAW.......142E}, in particular in motivating the
introduction of the cosmological constant in that paper.

As has already been mentioned, 
the philosophy of Mach, albeit firmly based in Newtonian physics,
was an important source of inspiration for Einstein. 
However, incorporating 
Machian ideas 
in a general relativistic cosmology 
presented serious difficulties.
After some early attempts had been discarded, Einstein in 
\cite{einstein:1917SPAW.......142E} adopted a spatially homogenous model of
the universe as a means of making a general relativistic cosmology compatible
with Machian ideas. 
Introducing a ``cosmological constant'' term $\Lambda
\ame_{\alpha\beta}$ in the
field equation of general relativity, 
which Einstein first motivated through a discussion of homogenous matter
distributions in Newtonian gravity, and assuming that there is a family of
observers who see the same matter density everywhere, 
led to a static universe filled with
a homogenous and isotropic matter distribution. 
The spacetime of the Einstein model is a Lorentzian cylinder. The line
element takes, up to a rescaling, the form 
$$
ds^2 = \ame_{\alpha\beta} dx^\alpha dx^\beta = - dt^2 + g_{S^3} \, . 
$$
This give a solution to \eqref{eq:EFE} 
with positive $\Lambda$, and with matter consisting of
a pressureless fluid with everywhere constant energy density. 

\begin{wrapfigure}{O}{0.5\textwidth}
\centering
\includegraphics[width=0.48\textwidth]{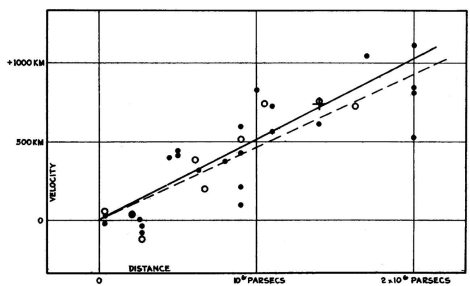}
\caption{Hubble's original 1929 graph \cite{hubble:1929PNAS...15..168Hb}} 
\label{fig:hub_1929}
\end{wrapfigure}

Shortly after Einstein's initial work on a static general
relativistic cosmology, 
Friedmann 
\cite{friedman:1922ZPhy...10..377F} proposed a model
of an expanding universe
\begin{equation}\label{eq:friedmann}
ds^2 = - dt^2 + a^2(t) g_\kappa
\end{equation} 
where $a(t)$ is a scale factor, $g_\kappa$ for $\kappa = +1, 0, -1$ is the
sperical, flat or hyperbolic metric. Line elements of the 
form \eqref{eq:friedmann} are also called Robertson-Walker line elements, 
see
below. 
During the 1920s,  Lema{\^{\i}}tre and Hubble showed, based on observational
work of
Slipher, Humason and others,  that redshift increases with
distance leading to the Hubble law, see figure \ref{fig:hub_1929}, which 
fits with the expanding Friedmann models. 
In the context of the expanding 
Friedmann models, Olbers' paradox can be resolved. Expanding Friedmann
models containing ordinary matter have $a \searrow 0$ at some time in the
past, where spacetime curvature and matter densities diverge.   
These 
models led, via the work of  Lema{\^{\i}}tre, Gamow, Hoyle and others,
to the hot big bang model which is the basis for the cosmological models in
use today.

E. A. Milne criticized the big bang models
on the basis that they introduced an extraneous
``cosmic time'' and also that they lacked explanatory
power (e.g. the sign of the spatial curvature is a priori undetermined). 
Instead, he proposed  
an extension of what he termed ``Einstein's cosmological principle'',
to the effect that ``The universe must appear the same to all
observers'' \cite{milne:1933ZA......6....1M}. Milne added to this the
postulate that observations are interpreted by each observer according to the
principles of special relativity and argued that this ``extended relativity
principle'' led to an essentially unique cosmological model. 

The
derivation of the general form of the line element compatible with the
isotropy of the universe, and also with Einstein's 
cosmological principle in the sense discussed by Milne
was given by Robertson 
\cite{robertson:1933RvMP....5...62R} 
and Walker \cite{walker:1935QJMat...6...81W} around the
same time, and found to be of the same form as that used by Friedmann and
Lema{\^{\i}}tre in their cosmological models. 
As pointed out by Robertson \cite{1933ZA......7..153R}, the 
general relativistic line element compatible with 
Milne's cosmology is a special case of \eqref{eq:friedmann}, 
namely the empty $\kappa=-1$ universe, which
is locally isometric to Minkowski space. This is therefore known as the Milne
model. 

It was a similar dissatisfaction with the lack of predictivity
 of general
relativistic cosmology that led Bondi, Gold and Hoyle
\cite{bondi:gold:1948MNRAS.108..252B, hoyle:1948MNRAS.108..372H} 
to introduce the
``perfect cosmological principle'', which is essentially a version of the
postulate of Milne, but viewed from the perspective of general relativity. By
allowing for creation of matter, they showed that it is possible to construct
an expanding cosmological model satisfying this principle. 
However, the perfect
cosmological principle tightly constrains the possible models of the universe
and the resulting steady state model 
is considered to be incompatible with observations. The book of Kragh 
\cite{kragh:1996cchd.book.....K} contains an interesting discussion of the
conflict between the steady state model and the now-standard 
``big bang'' cosmology. 

From the current perspective, it may be said that the introduction of
what Milne called Einstein's cosmological principle 
led to a class of general relativistic cosmological models. 
By introducing a collection of perfect fluids, a much simplified version of 
the problem of cosmological modelling reduces to the problem of fitting a
relatively small number of parameters to observational data, which could be
said to put cosmology on a similar footing as high energy particle physics. 
Indeed, as
mentioned by Peebles
\cite[Chapter I]{peebles:1993ppc..book.....P},
it was
Weinberg \cite{1972gcpa.book.....W} 
who introduced the notion, borrowed from high energy particle
physics, of a ``standard model'' into cosmology. 

At present, with the tremendous influx of data from observations of many
different types and at many different wavelengths, including observations of
the cosmic microwave background and galaxy surveys, it is often stated that
we are entering an era of precision cosmology. 
However, 
the widening 
  range of observational methods makes the process from observations  
to parameter estimation increasingly complex.
In particular, 
the prominent 
role of simplifying assumptions or principles in the formulation of cosmological
models and the model depence in the analysis of astronomical data, makes it 
important to keep in
mind the difference between a model which fits data to a high degree of
precision and a model which accurately describes the actual universe
\cite{peebles:2002astro.ph..8037P}.

\section{Cosmological models} 
For a Friedmann model, with line element of the form \eqref{eq:friedmann},
the stress energy tensor has the form 
$$
\aT_{\alpha\beta} = \rho \norm_\alpha \norm_\beta + p (\ame_{\alpha\beta} +
\norm_\alpha \norm_\beta) \, ,
$$
which is compatible with perfect fluid matter. Here $u^\alpha$ is the unit
timelike normal to the $t$ level sets, which in the special case of the
Friemann model coincides with the normalized 4-velocity of the fluid
particles, $\rho$ is the energy density of the
matter and $p$ is the pressure. 
We consider matter and radiation as described by a collection of fluids,
indexed by $i$, 
with linear equations of state, 
$$
p_i = \omega_i\rho_i .
$$
The Hubble constant (i.e. up to a constant factor the mean curvature of the $t$ level sets) is 
$$
H = \dot a/a
$$
In the special case of a Friedmann model, the contribution of the 
curvature of the $t$ level sets in the Einstein equations can be described in
terms of a fluid with equation of state $p=-\rho/3$, while the effect of the 
cosmological constant can be described by a fluid satisfying $p = - \rho$. Thus if we consider a 
simple model containing 
a fluid with pressure zero (dust), and with a cosmological
constant $\Lambda$, this 
can be described by introducing the dimensionless density parameters
\begin{align*} 
\Omega_m &= \frac{8\pi}{3H^2} \rho_m, & \text{``Matter'': } \omega = 0, \\
\Omega_\kappa &= - \frac{\kappa}{a^2  H^2},  & \text{``Curvature'': } 
\omega=-1/3, \\
\Omega_\Lambda &= \frac{8\pi}{3H^2}\rho_\Lambda,  & \text{``Vacuum'': }
\omega = -1 .  
\end{align*} 
The model can be 
parametrized by the present values 
$$
\Omega_{m0}, \Omega_{\kappa0},
  \Omega_{\Lambda 0} ,
$$
of the
density parameters. 
The conservation of matter 
and equation of state implies that the fluid densities $\rho_i$ depend only
on the scale factor 
\begin{equation}\label{eq:scalerel}
\rho_i \propto a^{-3(1+\omega_i)} .
\end{equation} 
The Hamiltonian constraint (i.e. the projection of the
Einstein equations \eqref{eq:EFE} on $u^\alpha$) takes the form 
\begin{equation}\label{eq:hamcon}
\Omega_m + \Omega_\kappa + \Omega_\Lambda = 1  ,
\end{equation} 
which, using \eqref{eq:scalerel},
can be written as 
\begin{equation}\label{eq:fried} 
\frac{H^2}{H_0^2} = \Omega_{0m} \left ( \frac{a_0}{a} \right )^3 +
\Omega_{0\Lambda} +
\Omega_{0\kappa} \left ( \frac{a_0}{a} \right )^2 .
\end{equation} 
Here $H_0, a_0$ are the present value of the Hubble constant and of the scale
factor respectively. Due to the uncertainty in the value of $H_0$, it is
usually given in terms of a dimensionless parameter $h$ as 
$$
H_0 = 100 \, h \text{km s$^{-1}$ Mpc$^{-1}$} .
$$
Equation 
\eqref{eq:fried} can be integrated to relate observable quantities, e.g. 
redshift and luminosity distance,  
for given values of the parameters $H_0,\Omega_{m 0}, \Omega_{\kappa 0},
\Omega_{\Lambda 0}$.

It is convenient
to study the global behavior of Friedmann models of the dimensionless density
parameters.  This analysis
is explained in \cite[Chapter 2]{wainwright:ellis:2005dsc..book.....W}, see
also \cite{lake:2005PhRvL..94t1102L,helbig:2012MNRAS.421..561H}. 
Due to the Hamiltonian constraint 
\eqref{eq:hamcon}, we have 
$\Omega_\kappa = 1 - \Omega_m - \Omega_\Lambda$. 
\newcommand{\bOmega}{\mathbf \Omega}
\newcommand{\Bb}{\text{\bf Bb}}
\newcommand{\dS}{\text{\bf dS}}
\newcommand{\Milne}{\text{\bf M}}
The fixed points of the
dynamial system in the $(\Omega_m, \Omega_\Lambda)$ plane are the 
Einstein-de Sitter big-bang model $\Bb = (1,0)$ 
and the spatially flat de Sitter model $\dS = (0,1)$, as well as the empty
$\kappa = -1$ Milne model $\Milne = (0, 0)$. One finds that 
$\Bb$ is a source and $\dS$ is a sink, while $\Milne$ is a saddle point. 
The static Einstein universe has $H = 0$, so the dimensionless parameters
$\Omega_m$ and $\Omega_\Lambda$ are 
ill-defined, but this point may be represented in an extended phase space 
as $\Ein = (\infty, \infty)$. This point is unstable, but 
is connected to the source $\Bb$ by an exceptional trajectory, which
separates the models which recollapse from those which expand forever. 

\begin{figure}[!hbt]
\psfrag{MAOMm0}{$\Omega_m=0$}
\psfrag{MAOMm1}{$\Omega_m=1$}
\psfrag{MAMpoint}{\Milne} 
\psfrag{MAFpoint}{\Bb} 
\psfrag{MAkm1}{$\kappa = -1$}
\psfrag{MAk0}{$\kappa = 0$}
\psfrag{MAkp1}{$\kappa = 1$}

\centerline{\includegraphics[height=1in]{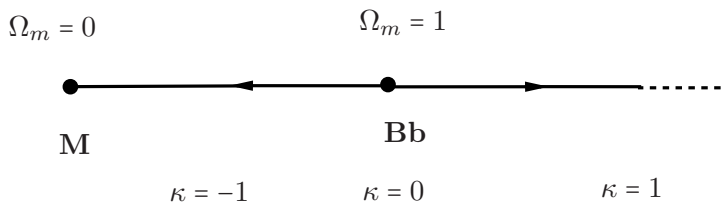}}

\caption{The dynamics of Friedmann dust models for $\Lambda = 0$} 
\label{fig:fried-dyn}
\end{figure}

Restricting to $\Lambda = 0$, the only fixed points are $\Bb$ and $\Milne$,
with $\Bb$ a source and $\Milne$ a sink, see figure \ref{fig:fried-dyn}.

\begin{figure}[!hbt]

\centerline{\includegraphics[height=3in]{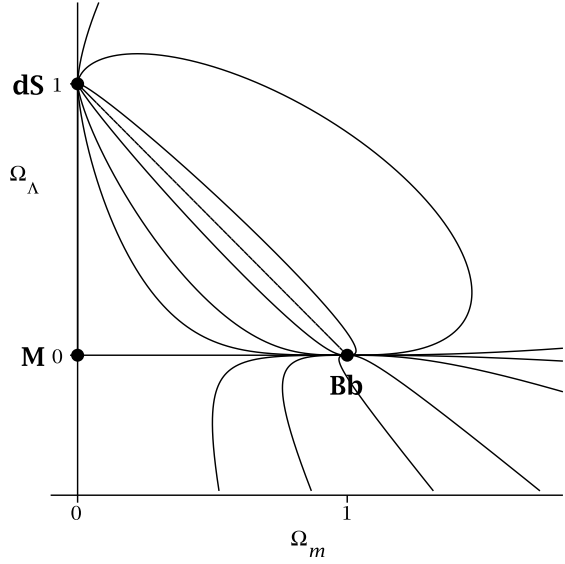}}
\caption{This figure shows some orbits for Friedmann cosmologies with dust and dark energy
  ($\Lambda$) in the $(\Omega_m, \Omega_\Lambda)$ plane. The Einstein-de
  Sitter point $\Bb=(1,0)$ is a source, the Milne point $\Milne = (0,0)$ is a saddle node, 
and the de Sitter point $\dS=(0,1)$
  is a sink. See \cite{lake:2005PhRvL..94t1102L} for background.} 
\label{fig:test}
\end{figure} 

The unstable Einstein-de Sitter universe 
$\Bb$
has slow volume growth $a \sim t^{2/3}$, while the stable 
Milne universe $\Milne$ 
has volume growth
$a \sim t$. In fact, this growth rate is maximal among $\Lambda = 0$ models.
This indicates that rapid volume growth goes together with
stability.

\begin{figure}[hbt]
\center
\includegraphics[width=0.48\textwidth]{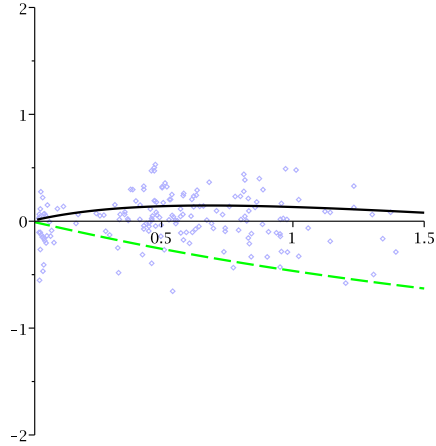}
\caption{Magnitude residual for SNe Ia Gold data 
  \cite{riess:etal:2007ApJ...659...98R} (dots) relative to the Milne model, 
plotted against redshift $z$. The black, solid curve is the standard model,
  while the green, dashed curve is the Einstein-de Sitter model. The
  horizontal axis is the Milne model. } 
 
\label{fig:SNIA} 
\end{figure}
Now we can give an extremely simplified description 
of the current situation in cosmology by saying that 
the laws of general relativity together with the cosmological
principle and observations leads to the ``standard model'' with the
cosmological parameters 
$$
\Omega_{\kappa0} \sim 0, \quad 
\Omega_{m0} \sim 0.3, \quad \Omega_{\Lambda0} \sim 0.7, \quad h \sim 0.7 .
$$
The standard model is a big bang model. There is an initial singularity, 
$a \searrow 0$ as $t \searrow 0$ and the universe expands indefinitely to the
future, $a  \nearrow \infty$ as $t \nearrow \infty$. 
The model predicts a hot big bang, which leads to the prediction of 
cosmic background radiation 
\cite{alpher:etal:1948PhRv...73..803A,alpher:herman:1948Natur.162..774A}. The observation of a highly
homogenous cosmic
background radiation with a spectrum close to that of a black body is a major
success of the big bang models of cosmology.

Most of the energy density in the standard model consists at present 
of as yet unknown ``dark matter'' (accounting for approximately 85\% of the
matter density) and ``dark energy'' in the form of the
cosmological constant. Dark matter, which for a long time has been broadly
accepted in astronomy and cosmolocy,
cf. \cite{vandenbergh:1999PASP..111..657V}, is distinguished from dark energy
by the
fact that its existence is motivated by studies of the dynamics of galaxy
clusters and galactic rotation curves, which are independent of the Friedmann
model which forms the basis of the standard model in cosmology. 
On the other hand,  
the cosmological
constant was deemed unacceptable on philosophical grounds
and entered the
standard model fairly recently, shortly before the year 2000; 
the effects of dark energy being seen only indirectly via cosmological
models and eg. studies of structure formation in the universe. 

The acceptance of $\Lambda$ 
came about only after the observation of the dimming of 
type Ia supernovae.
The
observations are interpreted as saying that the rate of expansion is
accelerated, i.e.  $\ddot a > 0$, which is incompatible with a Friedmann 
model filled with
ordinary matter and $\Lambda = 0$. Figure \ref{fig:SNIA} shows  
the supernova data compared to the standard model and Einstein-de
  Sitter. The horizontal axis is the Milne model
  ($\Omega_m=\Omega_\Lambda=0$).

\subsection{Cosmological problems} 

\begin{figure}[h!]
\includegraphics[width=0.48\textwidth]{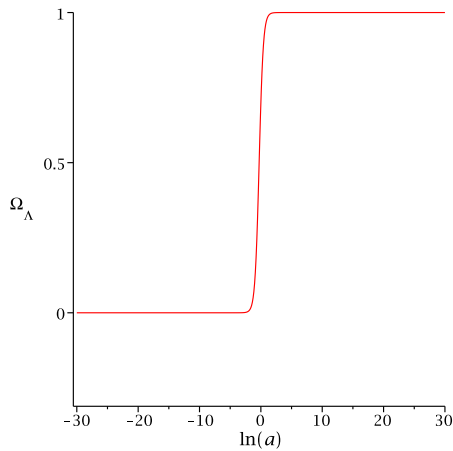}
\caption{The time evolution of the dark energy density, see
  \cite{carroll:2001LRR.....4....1C} for discussion.}
\label{fig:whynow} 
\end{figure}

One of the
important arguments 
against introducing the cosmological constant (apart from the difficulty of explaining the value
$\Lambda$ which appears motivated by cosmology from the point of view of
particle physics) 
has been
the \emph{coincidence problem}, which might also be termed the ``why now''
problem. 
Figure \ref{fig:whynow} shows
the time evolution of the dark energy density $\Omega_\Lambda$. We see that
it is only close to the present epoch that $\Omega_\Lambda$ becomes
significant, and in the 
later universe it will dominate the dynamics. 
Due to the different scaling behavior of the matter and
$\Lambda$ densities in view of \eqref{eq:scalerel}, the fact that these
are both of order unity at the present epoch is a coincidence that
could be argued to be contrary to the idea that we are not
``special observers''. 
In contrast, in the Einstein-de Sitter model 
the matter
density 
is time independent.

A related problem is the \emph{flatness problem}. Roughly speaking, this is
the question why $\Omega_\kappa \sim 0$ at present. In case $\Lambda = 0$
this can be seen to be problematic simply from figure \ref{fig:whynow}. Since
$\Bb$ is unstable, fine tuning of the initial conditions is required in order
to have $\Omega_\kappa \sim 0$ at present. 
Lake
\cite{lake:2005PhRvL..94t1102L} argues, using the presence of a conserved
quantity for the dynamics in the $(\Omega_m, \Omega_\Lambda)$-plane, that
fine tuning is not needed to have $\Omega_\kappa \sim 0$ throughout the
history of the universe. 

The universe is not exactly homogenous or isotropic; this holds at
best in an approximate sense on sufficiently large scales. 
This raises the problem of whether it is
possible to determine from observations, which are necessarily restricted to
our past light cone, to what extent, and at what scales, 
the assumption of homogeneity and isotropy 
is valid. 
\begin{figure}[!ht] 
\centering
\includegraphics[height=0.6\textwidth]{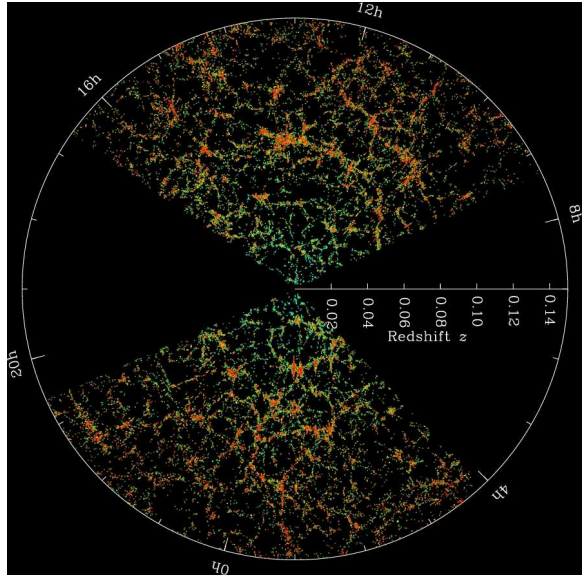}
\caption{Sloan Digital Sky Survey galaxy map, from \href{www.sdss.org}{www.sdss.org}} 
\end{figure} 
A problem here is that local isotropy (i.e. isotropy around the
world line of one observer) does not imply global homogeneity.

The Ehlers-Gehren-Sachs theorem gives conditions under which it is possible
to conclude from exact isotropy of the cosmic microwave background 
that the
universe is exactly isotropic. However, this result can fail in several
ways. 
For example, there are homogenous but
non-isotropic models where the CMB is exactly isotropic  at one instant in
time. Extensions of the EGS theorem to situations where only approximate
isotropy of the CMB holds are problematic, see
\cite{clarkson:2012CRPhy..13..682C,maartens:2011RSPTA.369.5115M,rasanen:2009PhRvD..79l3522R} and
references therein. 
This raises the problem of determining 
to what degree observations of the actual universe can be modelled and
analyzed in the framework of Friedmann models (and perturbations
thereof).  
One aspect of this
problem is the question whether there is a scale at which (statistical)
homogeneity and isotropy can be said to hold. Current estimates place this
scale at approximately 150$h^{-1}$ Mpc, see
e.g. \cite{marinoni:2012JCAP...10..036M}, see also
\cite{sylos-labini:2011CQGra..28p4003S}. 
However, recent observations
indicate the existence of inhomogenous structures of a dimension which may
be in conflict with isotropy at this scale, see 
\cite{clowes:etal:2012arXiv1211.6256C}. 
It is conceivable that observations which extend to
ever higher redshifts continue to yield evidence of structures in the
universe of a size comparable to the homogeneity scale. 
Some aspects of inhomogeneity in cosmology 
were recently surveyed in a focus issue of CQG, see
\cite{andersson:coley:2011CQGra..28p0301A} and references therein.

The question of how the potential effects of large scale inhomogeneities on
observations should be analyzed raises several important issues. Ellis has
formulated the ``fitting problem'', see \cite{2011CQGra..28p4001E} 
and references therein,  which asks about
the effect of analyzing observations from an inhomogenous universe via a
Friedmann model which is in some sense the ``best fit'' to the actual
universe. 
The effect on observations of the fact that the model universe used
to analyze data is only an approximation of the actual universe is sometimes
referred to as ``backreaction''. 
An important question here is whether perturbation
theory can be applied to take into account the deviation of the model from
the actual universe. 
Kolb and collaborators have argued
\cite{kolb:marra:matarrese:2010GReGr..42.1399K} that  
this analysis should take into
account the peculiar velocities due to the different expansion rate in the model and
the actual universe. 
Another effect of inhomogeneities which also sometimes
is referred to as backreaction, is the \emph{dynamical} effect of the
inhomogeneities on the expansion of the universe. A possible approach  is
to use averaging \cite{buchert:2011CQGra..28p4007B} or coarse-graining \cite{korzynski:2010CQGra..27j5015K}
to derive a set of effective equations modelling the universe. In order to
carry out such a scheme, one must introduce 
 closure relations which allow one to extract an autonomous
system. It is here worth mentioning the ideas on multi scale averaging, see
e.g. \cite{rasanen:2008JCAP...04..026R,wiegand:buchert:2010PhRvD..82b3523W}. In
particular, Wiltshire \cite{wiltshire:2007PhRvL..99y1101W}
argues that one should consider modifying the Copernican principle to take
into account the idea that we reside in a gravitationally bound structure in
a universe which has both bound systems and voids. 

It is apparent that the matter
distribution in the universe is ``lumpy'' due to the matter concentrations
in stars, galaxies and other structures, and inhomogenous due to the presence
of large scale voids and bound structures, and the effect of these must be
taken into account when analyzing observations, see Clarkson et al. 
\cite{clarkson:etal:2012MNRAS.426.1121C} for discussion. 
The optical properties of the
universe are, in the Friedmann models which form the basis for the standard
model of cosmology, calculated using the properties of a fluid which is used
to approximate the actual matter distribution. Thus it is necessary to
analyze whether the optical properties of a lumpy matter distribution differ
in a significant way from the optical properties of a fluid. Light from
distant stars passes through the gravitational wells of bound objects as well
as voids on the way to the observer, and the effect of this process must be
analyzed and compared to light passing through the fluid in a Friedmann
model. 
This problem has
been studied by among others Clifton et al
\cite{clifton:rosquist:tavakol:2012PhRvD..86d3506C}, see also
\cite{bentivegna:korzynski:2012CQGra..29p5007B}. In this context, we also
mention the so-called swiss cheese models, in which one attempts to analyze 
the optical effect
of voids and structure in the universe by introducing under-densities in a
background Friedmann model, see e.g. 
\cite{marra:kolb:matarrese:2008PhRvD..77b3003M} and references therein. The
swiss cheese models generally suffer from the limitation that the over-all
expansion of the model is determined by the chosen background Friedmann
geometry. 

In this situation one may contemplate introducing weaker cosmological
principles, incorporating ideas of statistical homogeneity, or weakening the
Copernican principle by restricting to matter bound observers as suggested by
Wiltshire.

As we have seen, the standard cosmological model is not located at 
a fixed point for the dynamical system governing the evolution of the
dimensionless parameters $\Omega_m, \Omega_\Lambda$,
rather it is close to the spatially flat orbit 
connecting the 
source $\Bb$ to the sink
$\dS$. Further, in that orbit,  $\Omega_m/\Omega_\Lambda$ takes on
all positive real values. Thus, we as observers are not in an asymptotic
regime, but rather, as mentioned above, at a special moment where
$\Omega_m$ and $\Omega_\Lambda$ are both of order unity. Thus, from this
point of view, we are neither in the ``early universe'' or the  
``late universe'' and we cannot argue that our current universe is singled
out as the asymptotic state of the evolution of the universe. 

This makes the situation in cosmology rather different from the situation in
many branches of physics where asymptotically stable objects are those which
one expects to find in nature. As an example, the Kerr black hole solution is
expected to be the unique stationary, asymptotically flat black hole
spacetime. In order to establish the astrophysical significance of this
solution, it is essential to prove that it is stable. This leads to the black
hole stability problem, one of the central open problems in general
relativity. The problem of determining from observations whether or
not for example the supermassive black holes expected to be found at the
center of most galaxies are Kerr black holes or not is being actively
studied. 

As was just mentioned, 
from the point of view of the current standard model in cosmology,
questions about the asymptotics of cosmological models do not appear to be 
the right ones
to ask. Nevertheless, such questions give 
rise to interesting mathematical problems
which we shall discuss in the rest of this paper. 
The questions about the asymptotic behavior of cosmological models include
the structure of the big-bang singularity and questions about the behavior in
the expanding direction. In particular we can ask: 
What does an observer in the late universe see?

\section{Asymptotics of cosmological models} 
\begin{wrapfigure}{O}{2.2in}
\psfrag{MAaslow}{$a(t) \sim t^{2/3}$}
\psfrag{MAacubed}{$a(t) \sim t$}
\includegraphics[width=2in]{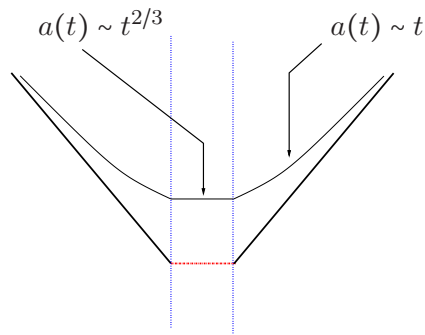}
\caption{A flat cosmological spacetime not isometric to Milne. A level
  set of cosmological time $t$ is shown. The vertical lines indicate
  the flat wedge which has been glued in.}
\label{fig:deformedmilne}\end{wrapfigure}

In this section we will describe a scenario for the asymptotic future 
behavior of
cosmological models with vanishing cosmological constant. 
Recall that the Milne model with line element 
$$
ds^2 = -dt^2 + t^2 g_{{\mathbb H}^3}
$$
where $g_{{\mathbb H}^3}$ 
is the hyperbolic 3-metric with sectional curvature $-1$, 
is isometric to 
the flat interior of the lightcone in Minkowski space. 
The Milne universe may
be viewed as the future of $O$, the origin in Minkowski space. This point
represents the big bang singularity in the Milne universe and is in the past
of all spacetime points (i.e. all observers). 
The cosmological time at a point $P$ is the proper time elapsed from the 
origin to $P$. The level surfaces of
cosmological time are simply the hyperboloids.
We next consider a flat, but non-isotropic model, which may be viewed as a
deformation of Milne. Let $I$ be a spacelike interval in Minkowski space and consider
the future of $I$. The resulting spacetime can be 
constructed by cutting the Milne spacetime by a timelike 
hyperplane through $O$ and gluing in a spacetime of the form 
$\Re^{2+1}
\times I$ with line element 
$$
-dt^2 + t^2 g_{{\mathbb H}^2} + dz^2 \, .
$$
The deformed Milne spacetime has a big-bang singularity given by the interval
$I$, and defining the cosmological time at $P$ as the maximal proper time of
any past inextendible geodesic starting at $P$ the level sets of
cosmological time are as in figure \ref{fig:deformedmilne}.

The deformed Milne universe is flat and empty,  
but not homogenous and isotropic.
Measuring the volume of co-moving regions in the deformed Milne universe we
see that in the deformed regions, the volume of the cosmic time levels grows
asymptotically 
as $t^{2/3}$, i.e. the growth rate of the Einstein-de Sitter universe, while
in the undeformed regions, the growth rate is asymptotically as $t$. The
behavior is similar for the level sets of the Hubble (mean curvature) time. 

In
particular, asymptotically as $t \nearrow \infty$, the volume fraction in the
undeformed region tends to $1$, while in the asymptotic past (near the big
bang) these regions have a negligible volume fraction. 

More general flat spacetimes may be constructed as the future of
sets (e.g. fractals) in Minkowski space, and quotients of these by the action
of discrete groups of isometries. Flat, or more generally
constant curvature spacetimes are examples of $G$-structures and such
spacetimes admitting compact Cauchy surfaces have been completely analyzed, 
starting with the work of Mess 
\cite{mess:const:curv}, see also \cite{MR2328922},
who analyzed the class of constant curvature 2+1
dimensional spacetimes admitting a compact Cauchy surface. For example, one
may show that 
the space of flat 2+1 dimensional spacetimes with Cauchy surface
of genus $g > 1$ is isomorphic to
$\partial \MM \times \MM$, where $\MM$ is 
Teichmuller space of surfaces of genus $g$ and $\partial \MM$ is the Thurston boundary. 
The particular case of 
constant curvature 
spacetimes with compact Cauchy
surface has been analyzed in \cite{andersson:etal:2012:CMCtime}. In
particular, it was shown there that such flat spacetimes can be globally
foliated by Cauchy surfaces of constant mean curvature (i.e. constant Hubble
time). 

The level sets of Hubble time can be related to the level sets of the
cosmological time by an application of a maximum principle, and one may show
that the volume growth of these level sets is comparable to that of the level
sets of the cosmological time. This leads to a generalization of the
statements made above for the simple deformed Milne universe, see
\cite{andflat}.  

\begin{figure}[!hbt]
\centering
\psfrag{MANeck}{Neck region -- slow volume growth}
\psfrag{MAHyperbolic}{Hyperbolic regions}
\includegraphics[width=0.6\textwidth]{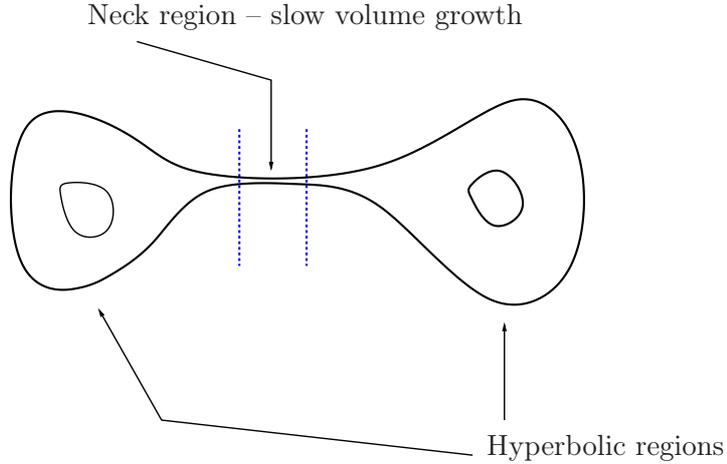}
\caption{Qualitative shape of Hubble level set} 
\label{fig:hublev}
\end{figure}

In view of the above mentioned work,  
these generalized Milne spacetimes may have a very complex (e.g.
fractal) big bang type initial singularity. In some cases their 
future asymptotics can be analyzed, see \cite{andersson:MR2216148}. 
One finds that the level sets of Hubble
time decompose into ``neck regions'' with slow volume growth, and
``hyperbolic regions'' with fast volume growth. The scale
free geometry 
of these level sets may
may be depicted as in figure \ref{fig:hublev}. 

In particular, one finds that in the asymptotically expanding direction, 
the volume fraction of asymptotically, hyperbolic (thick) 
regions dominate while the 
neck regions (thin) become insignificant.
Therefore, a ``typical'' (volume averaged) observer at late time
lives in a thick region. 

It is interesting to compare the relation between the thin and thick regions
to the overdense and void regions in an inhomogenous universe containing
matter, in particular in view of the fact that the thin regions have volume
growth approximating that of Einstein-de Sitter universe which has critical
matter density.

We now consider the generalization of the above picture to the case of
general, inhomogenous universes. We start by noting that the 
Lorentzian Einstein equations define a flow on the space of (scale free)
geometries. By analogy with the Ricci flow of Riemannian geometries, 
this may be termed the Einstein flow. 

\begin{figure}[ht]
\psfrag{MATeich}{$\MM$}
\psfrag{MAneginf}{$T=-\infty$}
\psfrag{MApinf}{$T=\infty$}
\centerline{\includegraphics[width=0.6\textwidth]{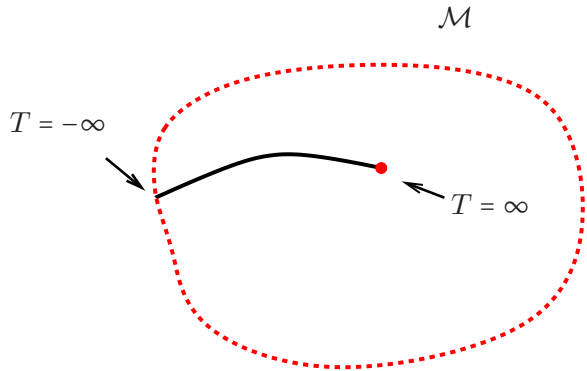} }
\caption{The Einstein flow in the 2+1 dimensional case} 
\label{fig:2+1}
\end{figure}

For simplicity, we consider spacetimes $(\aM,\ame_{ab})$ 
of dimension $D = d+1$ which are vacuum, i.e. with 
$$
\aR_{\alpha\beta} = 0 \, .
$$
Suppose $\aM$ admits a foliation by Cauchy surfaces
of constant mean curvature $H$.   
Introduce the dimensionless 
logarithmic constant mean curvature (Hubble) time
$
T=-\ln(H/H_0) ,
$
and consider the evolution of the scale free geometry $[g] = H^2 g$.
The Lorentzian Einstein equations define a flow
$
T \mapsto [g](T) ,
$
on the space of scale free geometries. 
In particular, in the 2+1 dimensional case, the 
Einstein equations correspond to a time dependent
Hamiltonian system on Teichm\"uller space \cite{andersson:etal:2+1grav},
and each universe corresponds to a curve connecting a point on the boundary
of Teichmuller space to an interior point, see figure \ref{fig:2+1}.

\begin{figure}[ht]
\centering
\includegraphics[width=0.4\textwidth]{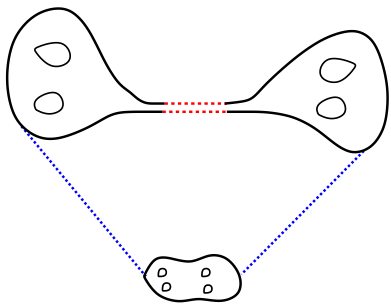}
\caption{The collapse of necks in the Einstein flow}
\label{fig:cusp} 
\end{figure}

One arrives at the following heuristic scenario 
\cite{fischer:moncrief:geometrization,anderson:geometrization:2001}.
Consider spacetimes with Cauchy surface $M$. The non-collapsing case
corresponds to the case where $M$ has negative Yamabe type. 
For $T \nearrow \infty$, $(M, [g])$
decomposes into hyperbolic pieces and Seyfert fibered pieces,
and this decompsition corresponds to a (weak) geometrization, cf. 
\cite{MR1888088}.
The Einstein flow in CMC time results in a thick/thin decomposition
  of $M$, where the thick (hyperbolic) pieces have full volume growth. As
  a consequence we have that 
in the far future, the hyperbolic pieces represent most
  of the volume of $M$, cf. figure \ref{fig:cusp}. 
Proving statements along the lines described above appears to be very
difficult, and one must therefore start by considering sub-problems. 

\section{Results on nonlinear stability} 
To give some perspective on the nonlinear stability problems introduced
above, we discuss some results on other stability problems in general
relativity. These are organized according to the asymptotic model spacetime.
The black hole stability problem, cf. \cite{2009arXiv0908.2265A} for
discussion and references, is not mentioned here. In the following, we
mention only the cases with conformally flat background spacetimes.

\subsection{Minkowski}
\begin{figure}[!h]
\centering
\psfrag{MAscrip}{$\Scri^+$}
\psfrag{MAi0}{$i_0$}
\psfrag{MAscrim}{$\Scri^-$}
\includegraphics[width=1.2in]{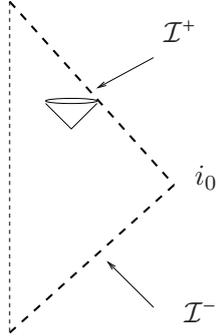}
\caption{Conformal diagram of Minkowski space}
\label{fig:mink} 
\end{figure}

First we consider the nonlinear stability of Minkowski space,
i.e. $\Re^4$ with line element  
$$
ds^2 = -dt^2 + dx^2 + dy^2 + dz^2.
$$ 
The conformal type of Minkowski space is
that of the Minkowski diamond, see figure \ref{fig:mink}. 
In this causal
diagram, each interior point represents a 2-sphere. 

Nonlinear stability holds, in the sense that 
for Cauchy data near Minkowski data, the maximal development is geodesically
complete and asymptotically Minkowskian. 
A key fact is that radiation carries  
energy through the conformal boundary $\Scri$. Due to the fact that the
nonlinearity in the Einstein equations is quadratic, it is necessary to
exploit a cancellation in the equations in order to prove stability. 

The first result in this direction is due to H. Friedrich 
\cite{friedrich:complete}, 
who proved that for data close to the data induced on a hyperboloid in
Minkowski space, one has nonlinear stability to the future, and with suitable
asymptotic regularity for the data, the maximal development has a regular
$\Scri^+$ to the future of the initial slice. 
The full nonlinear stability result was proved by Christodoulou and
Klainerman \cite{christo:klain}. 
This work was extended to include
the full peeling at $\Scri$ by Klainerman and Nicolo
\cite{2003CQGra..20.3215K}. 
A simpler proof of nonlinear stability, using wave coordinates (spacetime harmonic coordinates)
gauge was given by Lindblad and Rodnianski \cite{MR2680391}.
Using both of these methods, the proof of 
nonlinear stability can be readily adapted to the 
Einstein-matter system, provided that the matter fields do not destroy
the conformal properties of the Einstein equations. Examples include
a massless scalar field, which was included in the work of Lindblad and
Rodnianski, and a Maxwell field, see 
\cite{MR2531716}.

\subsection{de Sitter} \label{sec:dS}
Next we consider cosmological models with positive $\Lambda$. 
The canonical example is de Sitter space 
with line element 
\begin{equation*}
ds^2 = -dt^2 + \cosh^2(t)
g_{S^3} \,.
\end{equation*} 
This is conformal to a finite cylinder with spacelike conformal boundary, 
and hence one has 
future horizons and 
``locality'' at $\Scri^+$. Due to this fact, 
topology does not matter for the future dynamics 
(but
  cf. \cite{andersson:galloway}).
Due to the locality at  $\Scri^+$, we have that a suitable notion for
smallness in the stability argument can be defined locally in
space. We mention some results in this setting. 
H. Friedrich proved global nonlinear stability of de Sitter space for 
the Einstein-Yang-Mills system with positive cosmological constant
\cite{friedrich:glob:asympt}. 
H. Ringstr\"om proved a ``local in space'' small data global existence
results for the Einstein-$\Lambda$-scalar field system 
\cite{ringstrom:2008InMat.173..123R,ringstrom:2009CMaPh.290..155R}. 
The case of fluid matter was considered in this situation by
Rodnianski and Speck 
\cite{rodnianski:speck:2009arXiv0911.5501R} for the irrotational case, see
Speck \cite{speck:2011arXiv1102.1501S} for the Einstein-Euler system. 
Finally, the 
Einstein-$\Lambda$-Vlasov system has been studied by Ringstr\"om 
\cite{ringstrom:book:2012}.

\begin{figure}
\centering
\psfrag{MaScri}{$\mathcal I^+$}
  \begin{subfigure}[b]{0.5\textwidth}
    \centering
    \includegraphics[width=0.35\textwidth]{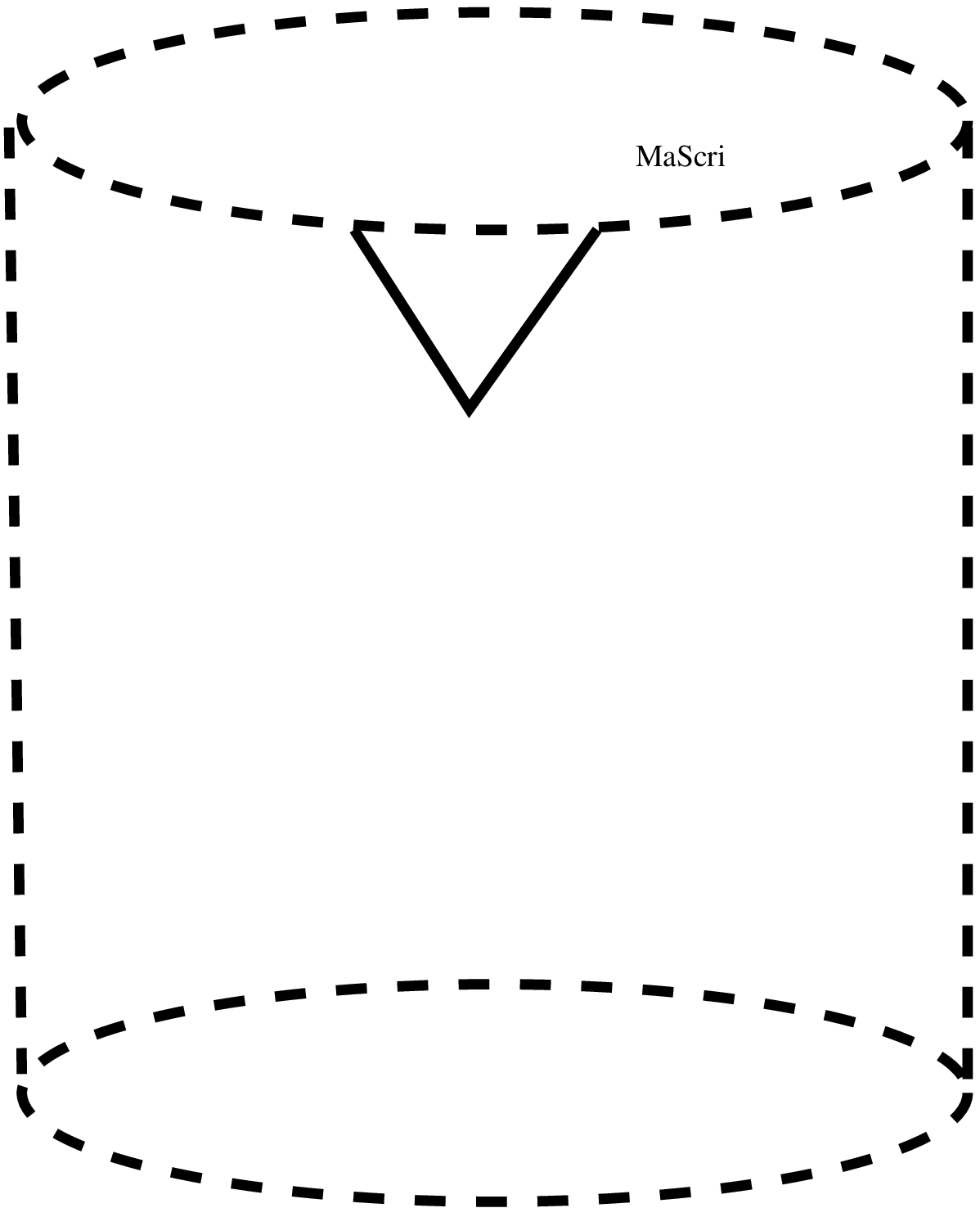}
\bigskip

    deSitter spacetime: conformal to a finite cylinder.
  \end{subfigure}%
  \begin{subfigure}[b]{0.5\textwidth}
     \centering
     \includegraphics[width=0.35\textwidth]{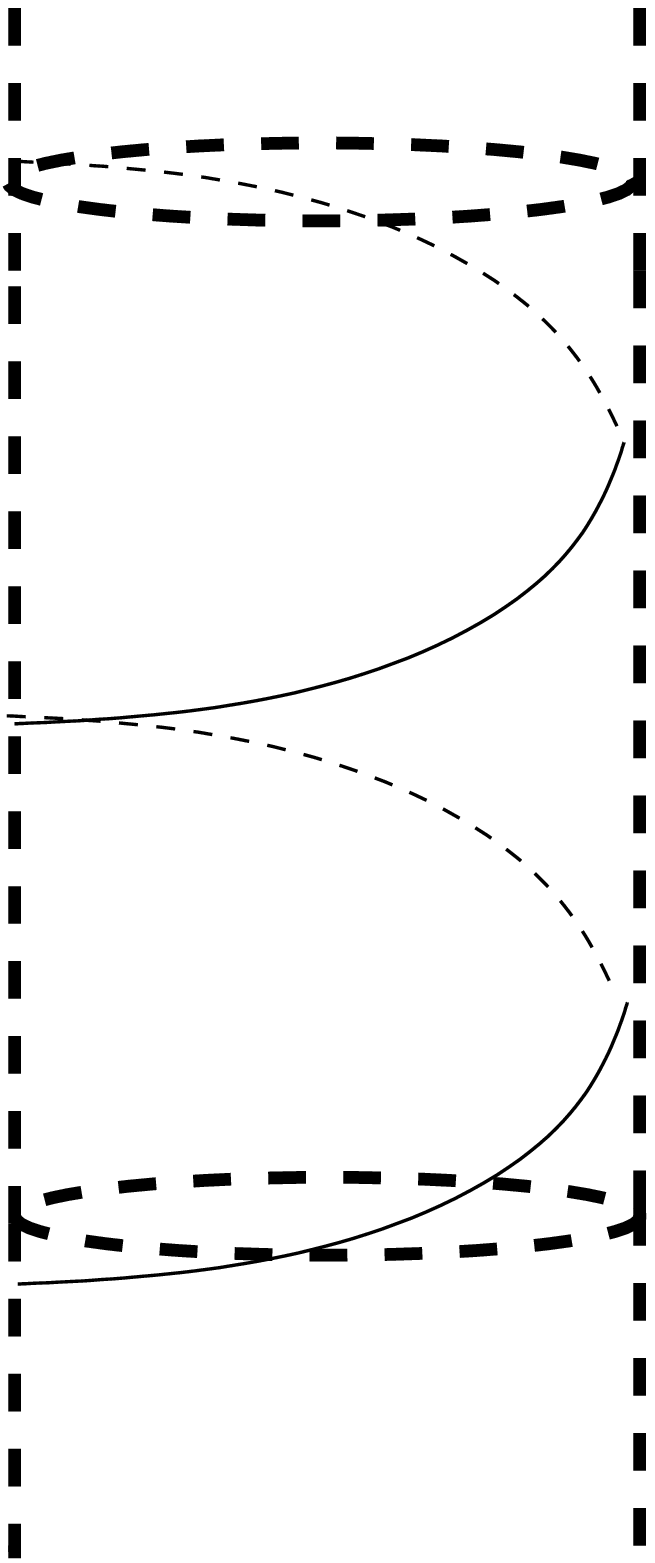}
\bigskip

    Milne spacetime: conformal to an infinite cylinder.
  \end{subfigure}
\end{figure}

\subsection{Milne} \label{sec:milne} 
Finally we consider the stability problem for a
cosmological models with $\Lambda = 0$. Here, the only general results are
for the vacuum case.  By
passing to a quotient of the Milne spacetime, we may consider a flat 
spacetime which has a Cauchy surface isometric to a compact hyperbolic
3-manifold. The line
element is 
$$
ds^2 = -dt^2 + t^2 g_{\HH^3}
$$
($\kappa = -1$ empty Friedmann) and the spacetime is conformal to an infinite
cylinder 
$$
- d\tau^2 + g_{\HH^3}
$$
In this case topology does matter, in the sense that an observer is
able to see the whole past of his spacetime. Since there is no future
conformal boundary, it is not possible to localize the future evolution
problem. 

Future stability for Milne with compact Cauchy surface as described above 
was proven by the author in collaboration with Moncrief for spacetime
dimension $d+1$, $d \geq 3$, cf. 
\cite{AMF,andersson:moncrief:2011:MR2863911}, see also 
\cite{reiris:2005:MR2708414, reiris:2010:MR2639547}. For the 2+1 dimensional
case, see \cite{andersson:etal:2+1grav}. 
Concerning the stability problem for the Einstein-matter systems in this
setting, much less is known than in the case with positive $\Lambda$. Some
sub-problems have been considered for the Einstein-Vlasov system in Bianchi
symmetry (spacetimes with a 3-dimensional Lie group acting by isometries on
Cauchy surfaces), see 
\cite{rendall:tod:1999CQGra..16.1705R},
\cite{heinzle:uggla:2006CQGra..23.3463H},
\cite{nungesser:2011JPhCS.314a2097N}.
Finally, we mention the work concerning test fluids on Friedmann backgrounds
by J. Speck \cite{speck:2012arXiv1201.1963S}.

The case of vacuum spacetimes with 
$U(1)$ symmetry leads after a Kaluza-Klein reduction to 2+1 dimesional
gravity with wave maps matter. The non-linear stability of the flat cones
over surfaces of genus $g > 1$ in
this setting has been studied by 
studied by Choquet-Bruhat and Moncrief, see \cite{ChB:moncrief:U(1),MR2098918}.

\section{Generalized Kasner spacetimes} \label{sec:genkas} 
In  section \ref{sec:milne} we discussed a stability theorem for the
future of a Cauchy surface in a class of spacetimes. The background spacetime
in that case is a Lorentz cone over a compact Einstein space with negative
scalar curvature, i.e. a
generalized Milne space. In particular these are warped products of the line
with an Einstein space. 
In this section we shall discuss a class of double
warped product spacetimes, with two scale factors. These spacetimes which
were considered in 
\cite{andersson:heinzle:MR2322531}
may be viewed as
generalized Kasner spacetimes. 
They have the form 
$$
\aM \cong \Re \times M \times N \, ,
$$ 
with
$(M,g)$, $(N,h)$, compact negative Einstein spaces of dimensions $m$, $n$,
respectively. The dimension of $\aM$ is $D = d+1 = m+n+1$.
We assume 
$\Ric_g = - (m+n-1) g$, $\Ric_h = -(m+n-1) h$.
and consider a line element on $\aM$ of the form 
$$
ds^2 = -dt^2 + a^2(t) g + b^2(t) h \,.
$$
Let $p = - \dot a/a, \quad q = - \dot b/b$, and introduce the scale invariant
variables 
$$
P = p/H, \quad Q = q/H, \quad A = \frac{1}{aH}, \quad B = \frac{1}{bH} \,.
$$
The  Einstein equations imply an autonomous system
for $(P,Q,A,B)$ with 2
constraints.
A dynamical systems analysis shows that the generic orbit has 
generalized Kasner behavior,  i.e.  $a \sim t^p$, $b \sim t^q$ 
at singularity, and is asymptotically Friedmann (in fact asymptotic to a
Lorentz cone spacetime) in the expanding
direction
$$
a,b =  t + O(t^{1-\lambda^*}), \quad \lambda^* > 0 \,.
$$
Friedmann is a stable node only in spacetime dimension 
$D \geq 11$.

\begin{figure}
\centering

\psfrag{F1}[cc][cc][1][0]{$(\text{F}_{1})$}
\psfrag{F2}[cc][cc][1][0]{$(\text{F}_{2})$}
\psfrag{FA}[lc][lc][1][0]{$(\text{F}_{\mathrm{A}})$}
\psfrag{FB}[rc][rc][1][0]{$(\text{F}_{\mathrm{B}})$}
\psfrag{B}[cc][cc][1][0]{$B$}
\psfrag{A0}[cc][cc][0.8][-62]{$A=0$}
\psfrag{B0}[cc][cc][0.8][-62]{$B=0$}

        \begin{subfigure}[b]{0.5\textwidth}
            \centering
            \includegraphics[width=0.35\textwidth]{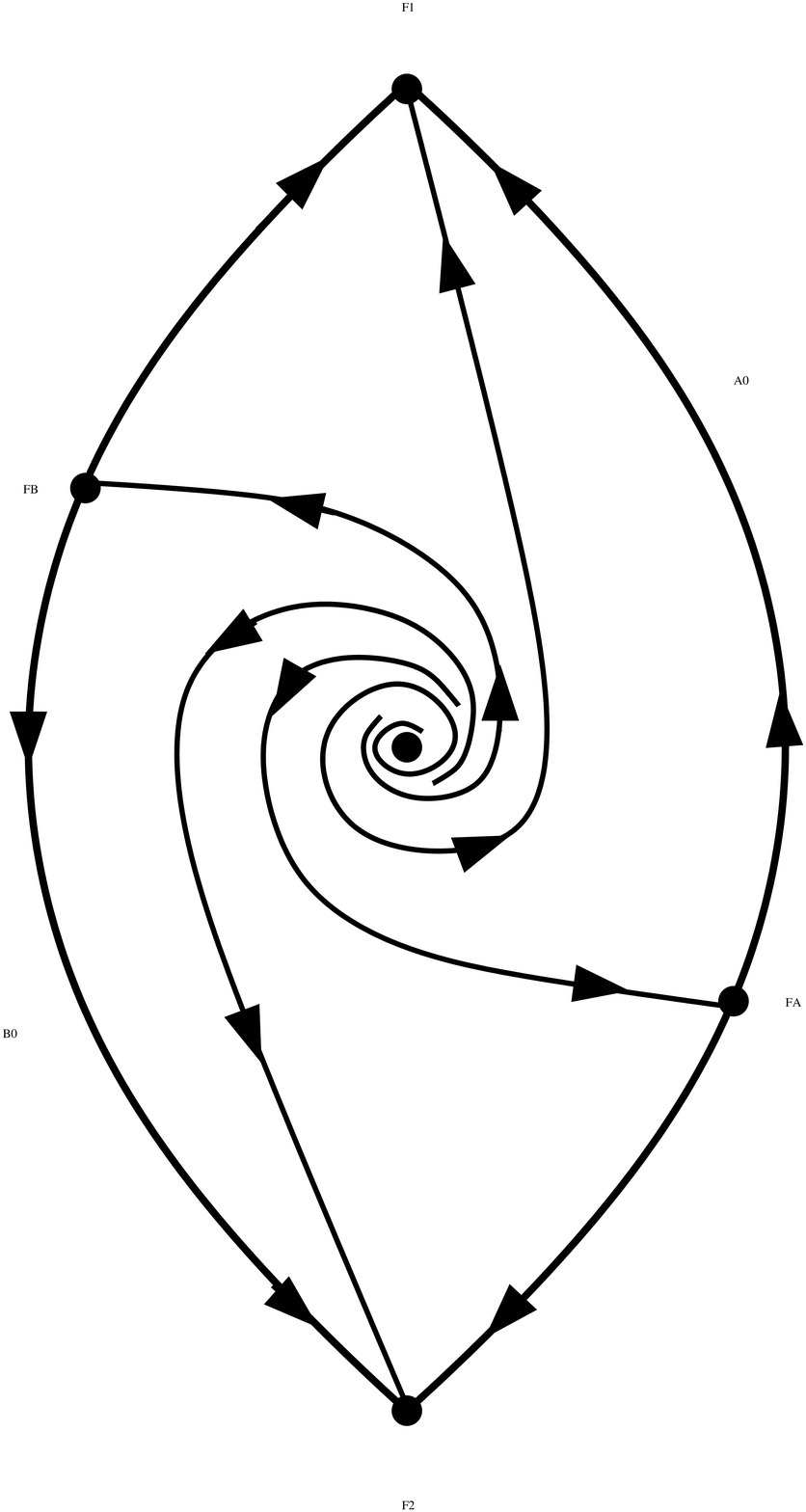}
            \caption{$D<10$}
            \label{fig:D<10}
        \end{subfigure}%
        \begin{subfigure}[b]{0.5\textwidth}
            \centering
            \includegraphics[width=0.35\textwidth]{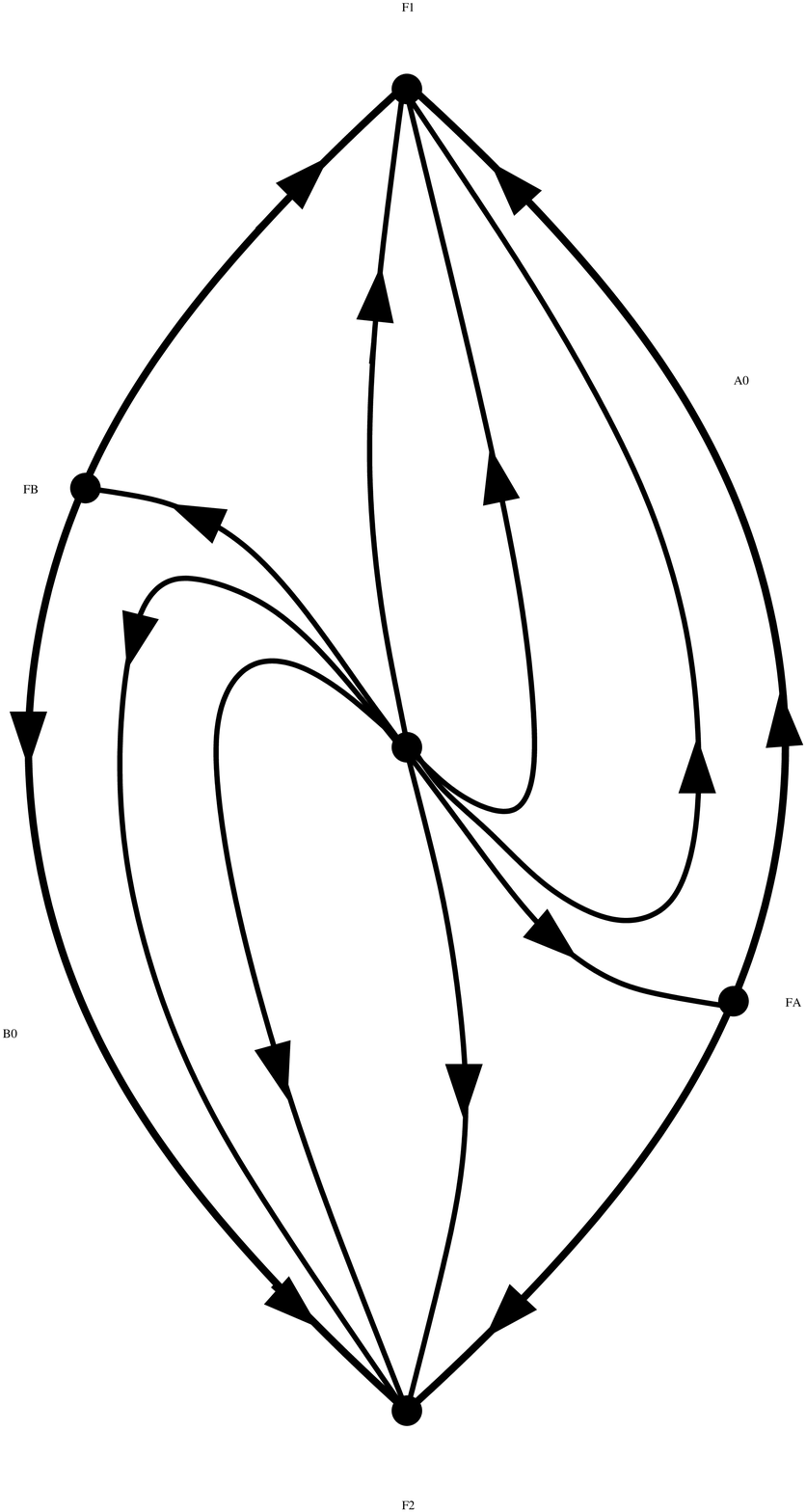}
            \centering
            \caption{$D>10$}
            \label{fig:D>10}
        \end{subfigure} 
        \caption{The dynamics of the generalized Kasner models \cite{andersson:heinzle:MR2322531}. The arrows
          point in the past direction. There are five fixed points, one of
          which is the Friedmann point in the interior of the phase space. 
          The Friedmann point  is
          a past unstable node for $D > 10$ and a unstable spiral point for $D <
          10$. The past stable fixed points $F_{1,2}$ satisfy 
          condition \eqref{eq:fuchscond} for $D > 10$. This implies quiescent
        behavior at the singularity for inhomogenous deformations of the generalized Kasner
        models in $D > 10$.}
\end{figure} 

\subsection{From $\alpha$ to $\omega$} \label{sec:ao} 
Belinski{\u\i}, Khalatnikov, and Lifshitz, \cite{BKL:1970AdPhy..19..525B} 
argued that a 
generic cosmological singularities in 3+1 dimensions in
spacetimes with ordinary matter 
is oscillatory. The picture developed by 
Belinski{\u\i}, Khalatnikov, and Lifshitz is often referred to as the BKL
proposal. BKL type behavior has been proved rigorously so far only for the
Bianchi VIII and IX models, see \cite{2000CQGra..17..713R}, where also strong
cosmic censorship for this class of models was shown. 
On the other hand, 
Belinski{\u\i} and Khalatnikov \cite{BK:1973JETP...36..591B} 
pointed out that 
cosmological singularities in 
spacetimes containing stiff fluid or scalar field
can be non-oscillatory, or \emph{quiescent}. 
The heuristic analysis of Belinski{\u\i} and Khalatnikov
was extended to the higher dimensional case by
Demaret et al. \cite{demaret:etal:1985} who showed that 
quiescent behavior at singularity in $D = d+1$ dimensions holds if the
condition 
\begin{equation}\label{eq:fuchscond}
1 + p_1 - p_d -
  p_{d-1} > 0
\end{equation}
holds, where  $p_a$ are the generalized Kasner exponents at the 
  singularity.
This heuristic analysis shows that (\ref{eq:fuchscond}) holds in vacuum
{\em only} if $D \geq 11$, and hence one expects that 
generic vacuum, $D < 11$ spacetimes have oscillatory singularity, while 
generic vacuum, $D \geq
11$ spacetime have  quiescent singularity.
It was shown in 
\cite[\S 4]{andersson:heinzle:MR2322531} that \eqref{eq:fuchscond}
holds for generalized Kasner spacetimes if $D \geq 11$, in agreement with the
result of  \cite{demaret:etal:1985}. 

As a step towards making this heuristic scenario rigorous, the author showed
with Rendall \cite{andersson:rendall:quiescent} that 
generic $D=4$ spacetime with scalar field  has quiescent singularity. In that
paper we constructed a full parameter family of Einstein-scalar field and
Einstein-stiff fluid spacetimes with quiescent singularity using Fuchsian
analysis. This work was extended to the case of $D \geq 11$ vacuum spacetimes
by Damour et al. 
\cite{damour:etal:2002STIN...0287809D}, again using a Fuchsian analysis.

One may use the techniques discussed above to prove that a type of 
global nonlinear stability holds for a class of generalized Kasner
spacetimes. 
It was shown in \cite{andersson:doublywarped} that for generalized Kasner
spacetimes as above, with $D \geq 11$, satisfying the additional condition
that the moduli space of negative Einstein metrics on $M,N$ is integrable
(which is expected to hold in general),  
there is a full-parameter family of
$C^\omega$ Cauchy data
on $M\times N$, such that the maximal Cauchy
development $(\aM, \ame)$ has a global CMC time function,
and has quiescent, crushing singularity. Further $(\aM, \ame)$ 
is future causally complete and is asymptotically Friedmann to the future,
with 
$g(T) \to \gamma_\infty^M + \gamma_\infty^N$, as $T \to \infty$, where 
$\gamma_\infty^M$ and $\gamma_\infty^N$ 
are negative Einstein metrics on $M,N$,
respectively. 
This applies to	 a large variety of factors $M,N$, and can easily
  be generalized to multiple factors.

\section{Concluding remarks} 
In this paper we have given brief overview of some of the ideas underlying
the general relativistic cosmological models which form the core of the
standard model of cosmology, and pointed out the need for an improved
analysis, both from the
physical and mathematical point of view, of 
the effect of deviations from homogeneity and isotropy in
the dynamics of cosmological models, and consequently in 
the analysis of cosmological data. Motivated by this, we have 
discussed some results on nonlinear stability for cosmological
models. We end by listing some open problems.

The exponential expansion caused by the presence of the cosmological constant
in the case $\Lambda > 0$ and also in the presence of certain
self-gravitating scalar field models for inflation makes the large data future
behavior of these models tractable and here there are several results which
do not require any symmetry assumtions, see section \ref{sec:dS}. 

For the
case $\Lambda = 0$ and ordinary matter, the situation is more delicate. 
The global behavior of cosmological models
is well understood in highly
symmetric cases, including the 3+1 dimensional 
Friedmann, Bianchi, Gowdy
(spatial $T^2$ symmetric, with symmetry action generated by hypersurface
orthogonal Killing fields) and so-called surface symmetric cases, see 
\cite{LA:MR2098914} and references therein. For the Bianchi case,
see the remarks in section \ref{sec:ao} and 
\cite{2001CQGra..18.3791R,heinzle:ringstrom:2009CQGra..26n5001H}, and for the
Gowdy case see \cite{ringstrom:2010LRR....13....2R} and references therein. 
However, for large
data, the asymptotic behavior of the general 
$T^2$, $U(1)$ (circle symmetric)  
and the full 3+1 case are
mostly open. 
Similarly, future stability is open in the 3+1 dimensional case 
for Einstein-matter models without symmetry
assumptions in the case $\Lambda = 0$. As an example, one would like to prove
nonlinear stability of Milne for Einstein-Vlasov. This is work in progress by
the author with D. Fajman. 

Our understanding of the behavior of cosmological models in the direction of
the initial singularity is also limited. 
The BKL proposal provides a
heuristic scenario which has been verified only in the Bianchi case, where
also strong cosmic censorship has been shown to hold, see
above. 
However, in spite of some recent progress
\cite{beguin:2010CQGra..27r5005B,liebscher:etal:2011CMaPh.305...59L,reiterer:trubowtz:2010arXiv1005.4908R}, 
even the question
whether the singularity in generic Bianchi models is local, is open. See 
\cite{heinzle:uggla:2009CQGra..26g5016H,heinzle:ringstrom:2009CQGra..26n5001H}
for references and discussion. For Gowdy models 
with $T^3$ Cauchy surface, Ringstr\"om
has proved that strong cosmic censorship holds, see
\cite{ringstrom:2010LRR....13....2R} for an overview, while for Gowdy with
Cauchy surfaces diffeomorphic to $S^3$ or $S^2\times S^1$, and the general
$T^2$ symmetric case (dropping the condition on hypersurface orthogonality)
the situation is much more complicated and cosmic censorship is open. 
In
particular, in the $T^2$ symmetric case, one has the new phenomenon of dynamical spikes, see 
\cite{andersson:etal:2005PhRvL..94e1101A,2012PhRvD..86j4049H}. 

The work by the author and Rendall, and by Damour et al. on quiescent
singularities, see section \ref{sec:ao} opens
up the problem of proving quiescent behavior at the singularity as well as
global nonlinear stability for an open set of Cauchy data (in a suitable
topology). This is work in progress by the author and Ringstr\"om. Work
on this type of stability problem for the Friedmann case was mentioned in a
recent talk by Speck  \cite{speck:oberwolfach-talk:2012}.
For the case $D < 11$ one may consider suitable Einstein-scalar
field models and for $D \geq 11$ one may formulate the global nonlinear
stability problem for the generalized Kasner backgrounds as discussed in
section \ref{sec:genkas}. Here it should be pointed out that the global
stability result mentioned there relies on Fuchsian methods and therefore
suffers from the same weakness as the work by the author and Rendall, and
Damour et al. on quiescent singularities. It would be interesting to
prove a true nonlinear stability result, stating that for an open set of
Cauchy data close to the generalized Kasner background data, the maximal
development is geodesically complete to the future, asymptotically
Friedmann, and with crushing singularity with geometry close to, in a
suitable sense, the singularity in the generalized Kasner spacetime. 

For the near future, I expect that 
numerical studies of cosmological models in GR, with 
less symmetry than the 2 Killing field models including LTB, $T^2$ and 
spherically symmetric models studied in detail so far,  
will play an important role in exploring the
future behavior of cosmological models. One can expect that 
such investigations will have an impact on 
both physical cosmology 
and the mathematical analysis of cosmological
models. 
\subsection*{Acknowledgements}
I would like to thank Ji{\v{r}}{\'{\i}} Bi{\v{c}}{\'a}k and
Bernd Schmidt for their comments on an early version of the paper.

\newcommand{\mnras}{Monthly Notices of the Royal Astronomical Society }
\newcommand{\prd}{Phys. Rev. D } 
\newcommand{\zap}{Zeitschrift f\"ur Angewandte Physik }
\newcommand{\pasp}{Publications of the Astronomical Society of the Pacific }
\newcommand{\sovast}{Soviet Astronomy }
\newcommand{\jcap}{Journal of Cosmology and Astroparticle Physics } 
\newcommand{\apj}{The Astrophysical Journal}

\end{document}